# Detecting bulk carbon ferromagnetism in graphene multi-edge structure

Chao Wang[1,3*], Nan Jian[2], Meijie Yin[2], Xi Zhang[2], Zhi Yang[1], Xiuhao Mo[1], Takashi Kikkawa[4], Shunsuke Daimon[4,5], Eiji Saitoh[4,6,7], Qian Li[8], Wensheng Yan[8], Dazhi Hou[9], Lei Yang[10], Dongfeng Diao[1,2,3*]

1. Institute of Nanosurface Science and Engineering, Shenzhen University, Shenzhen 518060, China;

2. Electron Microscopy Center of Shenzhen University, Shenzhen 518060, China;

3. National Key Laboratory of Radio Frequency Heterogeneous Integration, Shenzhen University, Shenzhen 518060, China;

4. Department of Applied Physics, The University of Tokyo, Tokyo 113-8656, Japan;

5. Quantum Materials and Applications Research Center, National Institutes for Quantum Science and Technology (QST), Tokyo 152-8550, Japan;

6. Institute for AI and Beyond, The University of Tokyo, Tokyo 113-8656, Japan;

7. WPI Advanced Institute for Materials Research, Tohoku University, Sendai 980-8577, Japan;

8. National Synchrotron Radiation Laboratory, and School of Nuclear Science and Technology, University of Science and Technology of China, Hefei, Anhui 230029, China;

9. International Centre for Quantum Design of Functional Materials (ICQD), Hefei National Laboratory for Physical Sciences at the Microscale (HFNL), and Synergetic Innovation Center of Quantum Information and Quantum Physics, University of Science and Technology of China, 230026 Hefei, China;

10. Key Laboratory of Education Ministry for Modern Design and Rotor-Bearing System, Xi'an Jiaotong University, Xi'an 710049, China.

E-mail address: dfdiao@szu.edu.cn.

**Abstract:**

The emergence of bulk carbon ferromagnetism is long-expected over years. At nanoscale, carbon ferromagnetism was detected in graphene nanoribbon segments[1–3], graphene triangulene dimers[4,5], and other graphene-like molecules[6–8] by analyzing their magnetic edge states via scanning tunneling microscopy(STM)[9], and its origin can be explained by local redistribution of electron wave function[3,5,8]. Recently, there are experimental efforts on discovering orbital ferromagnetism in twisted bilayer graphene[10,11], which is still limited in *STM* region. In larger scale, carbon ferromagnetism can be created by deliberately producing defects in graphite[12–14], and detected by macroscopic technical magnetization. Meanwhile, it becomes crucial to determine that the detected magnetization is originated from carbon rather than from magnetic impurities. One solution to eliminate this uncertainty is X-ray magnetic circular dichroism (XMCD). A contrast of dichroism at $\pi^*$ absorption energy was observed in the scanning x-ray microscope image of proton-irradiated graphite[15]. Nonetheless, a reproducible, full section of XMCD spectrum across C-1s absorption energy has not appeared yet, which should be decisive for assuring the indisputable existence of bulk carbon ferromagnetism. Besides, the lack of direct observation on the atomic structure of the ferromagnetic carbon leaves the structural origin of its ferromagnetism still in mist. In this work, for detecting bulk carbon ferromagnetism, we managed to grow all-carbon film consisting of vertically aligned graphene multi-edge (VGME), which wove into a three-dimensional hyperfine-porous network. Magnetization (M-H) curves and XMCD spectra co-confirmed bulk carbon ferromagnetism of VGME at room temperature, with the average unit magnetic momentum of ~0.0006 $\mu_B$/atom. The influence of magnetic impurities on magnetization was excluded by both absorption spectra and inductively coupled plasma mass spectrometry measurements. The spin transfer behavior also verified the long-range and robust feature of the bulk carbon ferromagnetism. Our work provides direct evidence of elementary resolved bulk carbon ferromagnetism at room temperature and clarifies its origin from π-electrons at graphene edges.

**Text:**

The emergence of bulk carbon ferromagnetism is long-expected over years. In nanoscale, there are theoretical underpinnings for carbon ferromagnetism such as the zigzag arrangement at graphene edges[1–3], vacancies[16], and adatoms[17,18], where the redistribution of electron wave function occurs and local ferromagnetism is generated. In experimental, local ferromagnetism was detected in graphene nanoribbon segments[1–3], graphene triangulene dimers[4,5], and other graphene-like molecules[6,7]. In addition, local magnetization in graphene was modulated by adatoms and doping[19,20]. Recently, there are experimental efforts on discovering orbital ferromagnetism in twisted bilayer graphene[11,21]. These cases of low-dimensional carbon ferromagnetism are at cluster scale with several hexagons in-plane, and a few layers out-of-plane. The ferromagnetic orders were detected under scanning tunneling microscopy by measuring local magnetization or electric polarization via analyzing their edge states.

Carbon ferromagnetism in larger scale can be created by deliberately producing defects in graphite through irradiation[12–14], oxidation[22], or mechanical exploiting[23]. This type of ferromagnetism is detectable in macroscopic technical magnetization. The hysteresis loop in magnetization curve is ascribed to magnetic moments of edge states at the defected positions. Meanwhile, it becomes crucial to determine that the detected magnetization is originated from carbon rather than from magnetic impurities, because carbon ferromagnetic signals are several orders of magnitude lower than magnetic element, and M-H curve has no elemental resolution. Just out of this fact, the experimental observations of macroscopic ferromagnetism in carbon materials were always challenged on the issue of impurities, or at least not well recognized, since it is difficult to rule out the influence of impurities. There are even reports against carbon ferromagnetism after careful experiments[24,25].

One solution to eliminate this uncertainty is X-ray magnetic circular dichroism (XMCD), which can recognize carbon ferromagnetism from the change of its absorption spectra under opposite magnetic field. A contrast of dichroism at 284 eV was observed in scanning X-ray microscope image of proton-irradiated graphite[15]. Nonetheless, a reproducible, full section of XMCD spectrum across C-1s absorption energy has not appeared yet, which should be decisive for assuring the indisputable existence of bulk carbon ferromagnetism.

For ferromagnetic carbon in any form, the atomic structure is the source of its unique electronic structure that produces ferromagnetic order. In the studies of low-dimensional ferromagnetic carbon allotropes, the visualization of their atomic structures significantly improved the interpretation on the origin of local carbon ferromagnetism. However, in the case of bulk ferromagnetic carbon, the lack of direct observation on the atomic structure leaves the structural origin of its ferromagnetism still in mist.

In this work, we managed to grow all-carbon film consisting of vertically aligned graphene edges, which wove into a three-dimensional hyperfine-porous network. The vertical graphene multi-edge (VGME) structural feature was identified by high resolution transmission electron microscopy (HRTEM). M-H curves and XMCD spectra co-confirmed the room temperature ferromagnetism of VGME. Carefully repeated measurements were carried out to reduce the noise, and the average unit magnetic momentum of ~0.0006 $\mu_B$/atom was determined. The influence of magnetic impurities on magnetization was excluded by absorption spectra and inductively coupled plasma mass spectrometry measurements. Angular dependent magnetoresistance of VGME was measured in room temperature, and the spin transfer behavior verified the long-range and robust feature of ferromagnetism. Our work provides direct evidence of elementary resolved bulk carbon ferromagnetism at room temperature and clarifies its origin from $\pi$-electrons at graphene edges.

The VGME structure was produced in electron cyclotron resonance (ECR) plasma via a physical vapor deposition (PVD) approach with a low-energy-electron-irradiation technique, which was explained in detail in our previous work[26,27]. This approach allows us to produce VGME structure at wafer scale on variable substrate surfaces include but not limited to $SiO_2$. Fig. 1a-c depict a sequence of plan-view high-angle annular dark-field (HAADF) images of the as deposited film consist of VGME structure (VGME film), progressively zooming in. Fig. 1a reveals the porous architecture of the VGME film as black void regions are distributed throughout the film. The pores are of irregular shapes, and their average size generally does not exceed 2 nm according to statistical measurements (see Fig. S1). Bright regions of the images represent the solid part of VGME structure, which consists of turbostratic graphene. The binding configuration of the turbostratic graphene was confirmed by Raman and XPS spectra (see Fig. S2). The curving and crosslinking nature of the graphene layers can be clearly observed by the line-shaped projections in Fig.1b. The dangled edges around superfine pores are highlighted in purple dashed lines. Theses vertically dangled graphene edges are believed crucial for the emergence of ferromagnetic moment. Fig.1c magnifies the graphene layers around one of the pores. The interlayer spacing of 0.35 nm is revealed as labeled by black arrows in the image, corresponding to the van der Waals distance between graphene layers. Inside each graphene layer exists the spacing of 0.21 nm which corresponding to the distance between neighboring zigzag edges, as labeled by yellow arrows and illustrated by the inset sketch. According to the viewing direction and the appearance of the zigzag edge projection, it is verified that the zigzag direction of the graphene layer is in parallel with viewing direction, i.e., vertical to the substrate surface. The side view TEM image of the VGME film in Fig. 1d confirms that most graphene layers orient vertically to the film/substrate interface. Fig. 1e further highlights the features of crosslinking, splitting, and merging between neighboring graphene layers by colored dash lines. Such features suggesting the observed pores are not perpendicular linear tunnel through the thickness of the film, instead, they are discontinuously distributed inside the whole film. The blue square part of the image is magnified in Fig. 1f, in which a line-shaped projection of the graphene layer is observed with the inner spacing of 0.12 nm, corresponding to the spacing between two nearest atomic planes which are along armchair direction, as labeled by yellow lines and illustrated by the inset sketch. The spacing of 0.12 nm in side view co-demonstrate that the vertically aligned edges are zigzag type, as Fig. 1c does. This structure thereby is named as vertical graphene multi-edge (VGME) structure, which consist of multiple graphene layers aligned parallel to the film grown direction. These graphene layers further wove into a three-dimensional hyperfine-porous network.

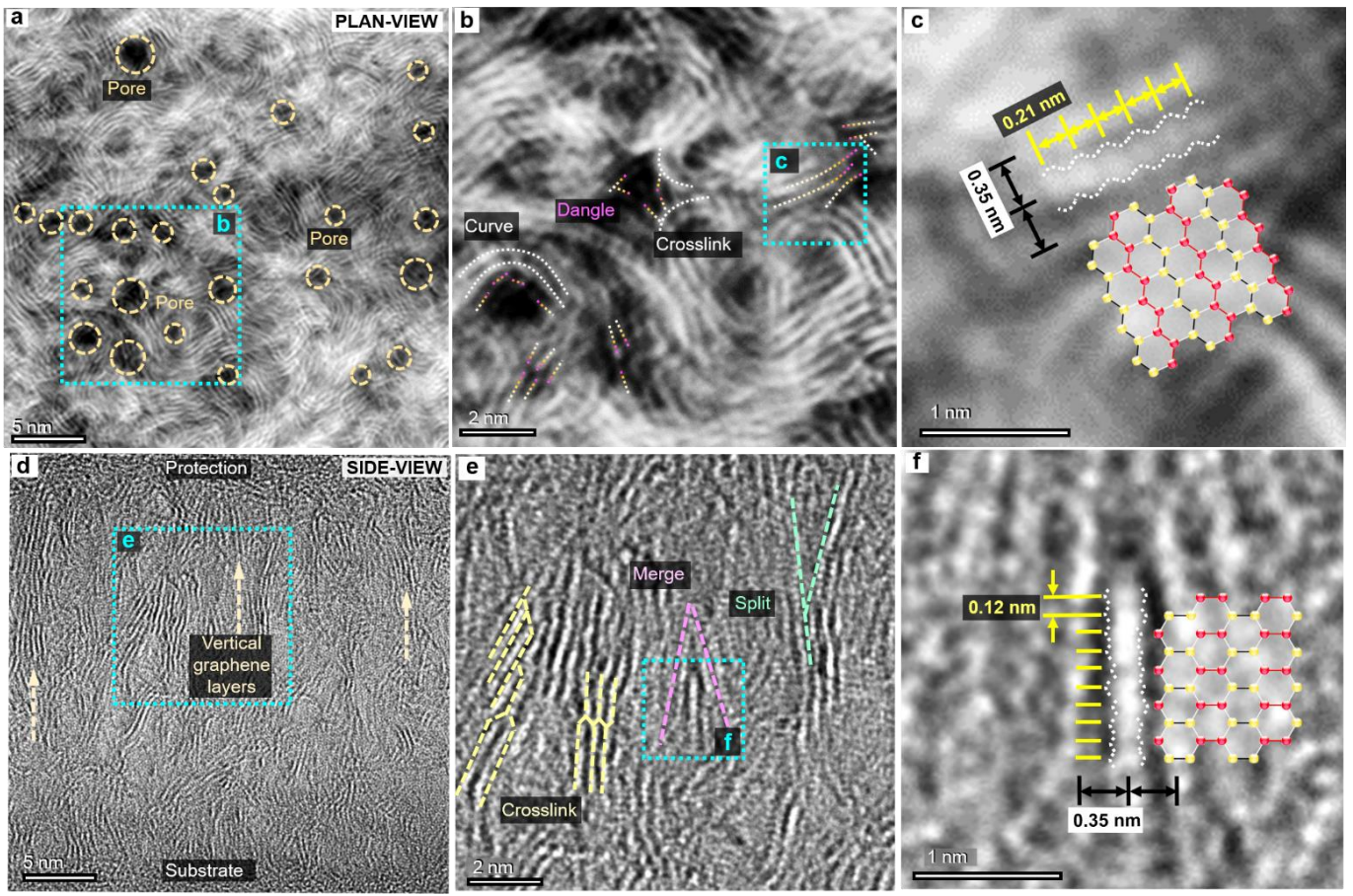

**Fig. 1 Atomic image of vertical graphene multi-edge (VGME) structure. a,** Plan view HADDF image in which the superfine pores in the structure are observed, as highlighted in the image by dashed circles. The bright lines in the figure represents the projections of curved graphene layers. A square area b is selected for zoom-in observation. **b,** zoom-in image of the selected area in a. The curve, dangle, and crosslink features of the graphene layers can be observed clearly, as guided by dashed lines. A square area c is selected for further zoom-in observation. **c,** zoom-in image of the selected area in b. The atomic structural detail of the graphene layer can be identified, as indicated by dashed lines. The spacing of 0.35 nm between graphene layers and 0.21 nm between zigzag edges are labeled. **d-f,** Side view TEM images of the VGME film, progressively zooming in. The vertical orientation of the graphene layers can be seen in d; the crosslink, merge and split features can be seen in e; the spacing of 0.35 nm between graphene layers and 0.12 nm between nearest atom lines along armchair direction are labeled.

The magnetization (M-H) curve of the above VGME structure is shown in Fig. 2a in blue color. The curve was measured at the temperature of 300 K, and the saturation magnetic field H of 0.7 T. The direction of magnetization is vertical to the VGME film, that is parallel to the graphene edges. A hysteresis loop with saturation feature is clearly shown in the M-H curve. The magnetization $M_s$ at 0.7 T, the residue magnetization at 0 T $M_R$, and the coercive magnetic field $H_c$ (at 0 magnetization) are listed inside the figure. The $M_s$ value of VGME film at 0.7 T is 4.7 emu/g, converting to unit magnetization of ~0.0008 $\mu B$ per atom. This is three order of magnitude higher than reported in defected HOPG[23]. Furthermore, the M-H curve shows rapid increment at low field, and reaches 4.3 emu/g at ±0.2 T. When magnetic field further increases, the magnetization slowly saturates. For the need of compare study, we prepared an amorphous carbon film (a-C) with the same technique, during which the irradiating energy of electrons were at a rather low level so that VGME structure was not formed. Comparing measurement results of HOPG and a-C samples are also given in Fig. 2a in grey and yellow curves. These curves are magnified for better resolution in Fig. 2b and c. The $M_S$ value of HOPG sample is $1.1\times10^{-3}$ emu/g and the $H_c$ value is 25 Oe, close to the earlier reported value [23], implying that the results of our hysteresis measurement are reasonable. The $M_s$ value of a-C film is $1.4\times10^{-4}$ emu/g, four orders of magnitude lower than that of VGME structure. Since the VGME and a-C samples are prepared by the same technique, in can be inferred that potential contribution to $M_s$ from impurities could not exceed $1.4\times10^{-4}$ emu/g. Therefore, the detected hysteresis loop and large saturation magnetization of VGME structure could be ascribed to bulk carbon ferromagnetism. To verify this interpretation, XMCD measurements were introduced for elemental-resolved magnetic detection.

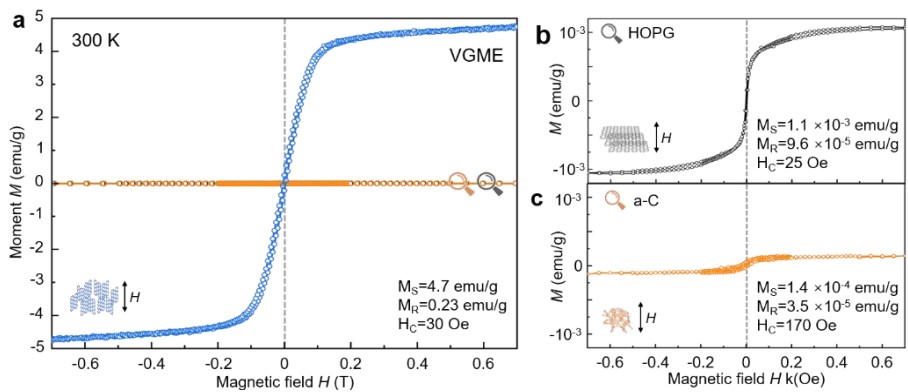

**Fig. 2 Magnetization curves of VGME, HOPG and a-C structures at 300 K. a,** the hysteresis loop of VGME with magnetic

field H perpendicular to film surface, as indicated inside the figure. The saturation magnetization $M_s$, residue Magnetization $M_R$ and coercive field $H_c$ derived from the loop are listed. The $M_s$ reaches 4.7 emu/g, equals to 0.0008 $\mu_B$/atom. M-H curves of and a-C are presented in the same scale in the figure. **b,** magnified hysteresis loop of HOPG with H perpendicular to graphene layer, as indicated inside the figure. **c,** magnified hysteresis loop of a-C with H perpendicular to the film surface.

Fig.3a summarizes the X-ray absorption near edge spectra (XANES) collected from VGME and a-C film as well as HOPG in vertical and horizontal positioning directions. The near-edge detailed spectra are highlighted in yellow background, and the X-ray inject direction are presented above each spectrum. All samples show clear C-1s K-edge absorption signals. The XANES of the VGME sample shows a strong peak at $\pi^*$ position of ~284.5 eV, which is more similar to that of vertical HOPG than to horizontal HOPG, suggesting the strong X-ray absorption from 2p orbital. This is due to the reason that the incident direction of the X-ray was perpendicular to the film surface, and the graphene layers in VGME are vertically grown as were observed in TEM images. It should be pointed out that the $\pi^*$ peak from VGME sample is extremely stronger than from the other three. Such feature can be ascribed to the vertical graphene multi-edge structure, which produces abundant edge states those are more active under X-ray excitation. In order to check whether our results were affected by magnetic impurities, the absorption spectra at the energy ranges of L-edge of Fe, Co, and Ni were acquired, and no signals were detected, as can be seen in Fig 2b-d. This indicates that the observed magnetization signal of the sample does not originate from magnetic impurities other than carbon. In addition, the trace of metallic impurities in VGME film were analyzed by inductively coupled plasma mass spectrometry (ICP-MS), and the results showed that the total content of metallic impurities (mainly Al, Cr and Fe) are less than 30 ppm (see Fig. S3). Such content level coincides with the above interpretation. Therefore, the influence of metallic impurities on the electronic structure of total VGME sample, for example, via alloying or proximity effect, can be neglected.

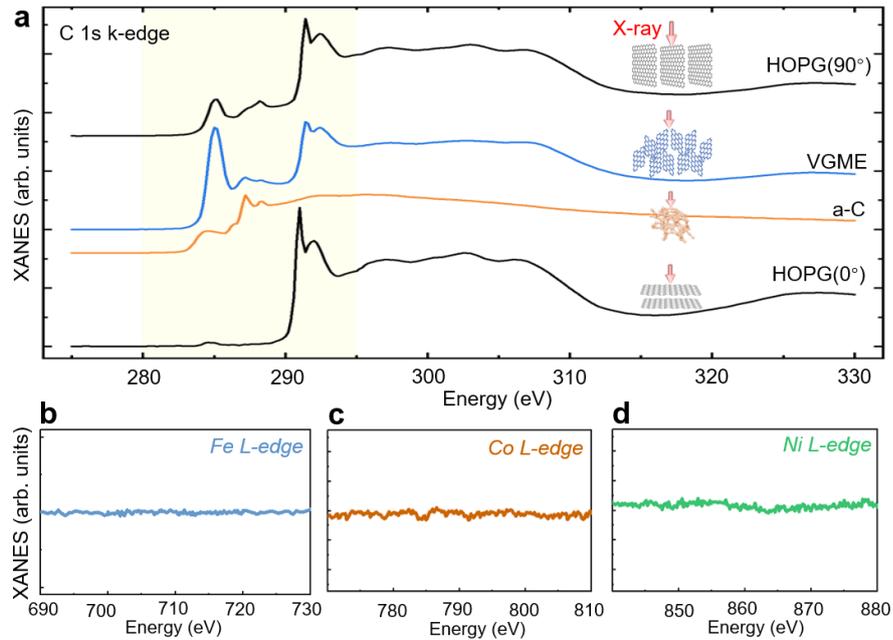

**Fig.3 X-ray absorption spectroscopy of the VGME, HOPG and a-C structures. a,** C-1s K-edge absorption spectra of VGME, a-C and HOPG (vertical and horizontal) samples. The near-edge detailed spectra are highlighted in yellow background, and the X-ray incident direction are presented beside each spectrum. **b-d,** absorption spectra of VGME sample at the energy ranges of L-edge of Fe, Co, and Ni, respectively. No signal is observed, suggesting the influence of magnetic impurities can be neglected.

Fig. 4 compares the XMCD spectra of VGME and a-C structure, which are detected at 300 K with the external, out-of-plain magnetic field of 0.7 T. In order to guarantee the reproducibility and increase signal-to-noise ratio, the spectra were repeatedly taken for 10 times, and the average signals from increasing scanning times are exhibited in Fig 4a-f. As shown in the figures, the dichroism of $\pi^*$ peak at ~284.5 eV becomes more profound after multiple repeating measurements on VGME sample, and the noise level is effectively reduced, so that the dichroism peak can be identified at rather low amplitude. The profound feature of the dichroism at $\pi^*$ excitation energy of VGME structure provides a direct evidence of bulk carbon ferromagnetism originated from 2p orbital. Since the dichroism appears at the same energy as in XANES spectra, its origin can be ascribed to the edge states in the VGME structure, as mentioned above. The amplitude of VGME-XMCD peak is ~0.005 and stabled after repeatedly measurements, corresponding to the effective magnetic moment of ~0.0006 $\mu B$ per atom by relative area ratio method (see description in method section). This value is close to its $M_s$ that calculated from the hysteresis loop. On the other hand, after 10 times of repeating measurements, the XMCD spectrum of a-C film only shows fluctuations at the amplitude of ~0.002, such amplitude is comparable to background noise level of the measurements, as indicated in each panel with parallel dashed lines. We note that for the L-edge absorption spectra of transition metals, spin and orbital angular momentum can be separated by introducing sum rule[28]. Yet so far, this methodology is not suitable for the K-edge absorption spectra as in the case of carbon. Recent breakthrough in the analysis of K-edge spectra shed light on extracting the orbital contribution to the s-p dichroism signal[29]. Therefore, we anticipate the XMCD of C-1s K-edge to unveil more information about the origin of bulk carbon ferromagnetism and benefits deeper understanding of carbon electronic structures.

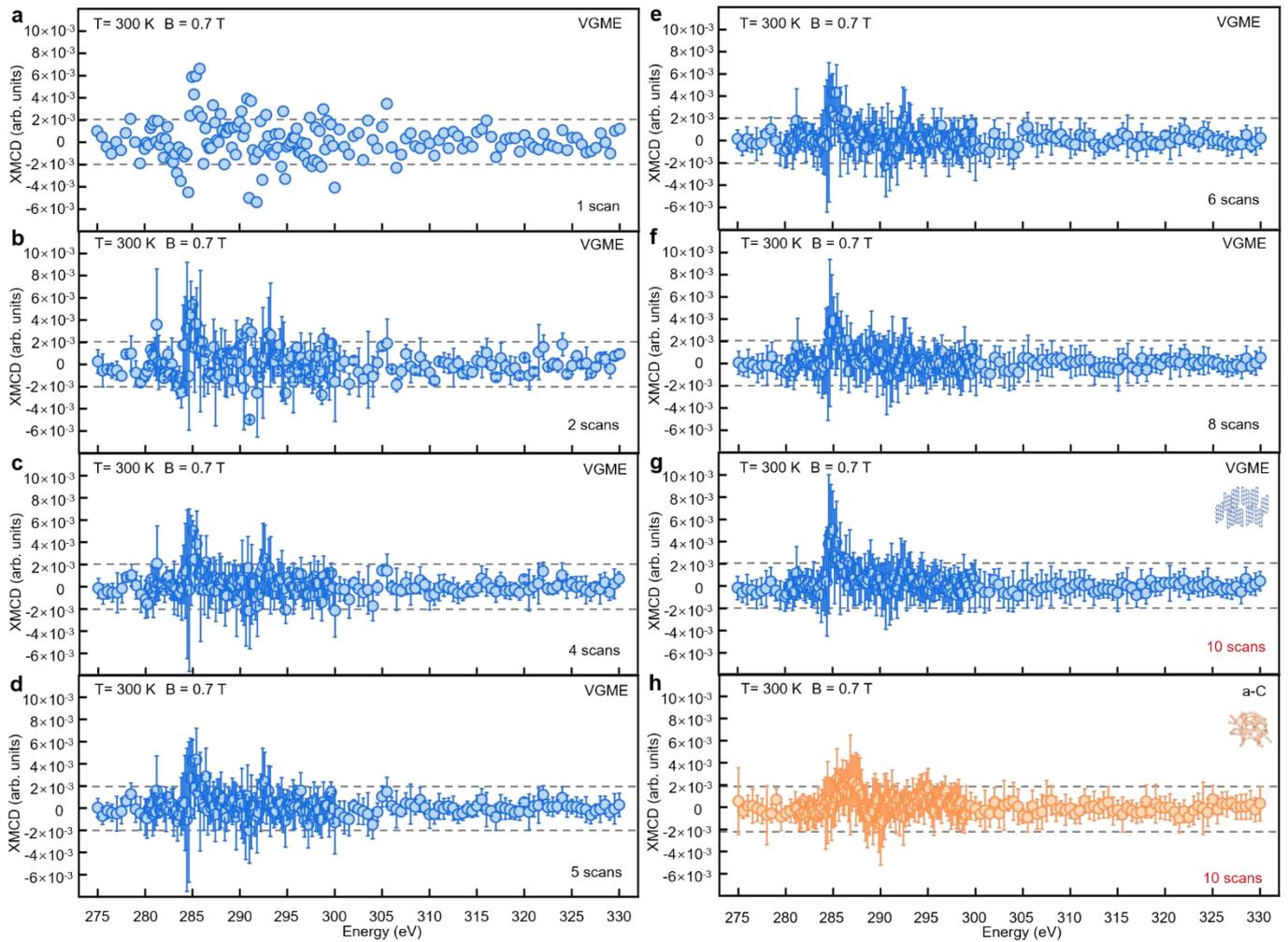

**Fig. 4 XMCD spectra of VGME and a-C structures detected at 300 K with the external out-of-plane magnetic field of 0.7 T. a-g**, the average results of XMCD spectra of VGME structure with increasing scan times from 1 to 10 scans. The

dirhosim peaks at ~284.5 eV are indicated by solid curves in each spectrum. **h,** the average result XMCD spectra of a-C structure after 10 scans. The noise ranges between ±0.002 a.u. are indicated by parallel dashed lines in each panel.

For a bulk material with long-range ferromagnetic order, one should expect multiple magnetic dynamic behaviors other than a classic hysteresis loop. For such purpose, we investigated the spin transport behavior in the VGME film. Fig. 5a-b illustrate the principle of the measurement setup. A platinum layer Pt is coated on top of the VGME or a-C film to build a double layer junction, being referred as Pt/VGME or Pt/a-C. When an electric current $J_c$ is injected through Pt layer along X-axis, a spin current $J_s$ is naturally formed along Z-axis due to spin-Hall effect (SHE), in which the spin polarized electrons diffuse to the interface of the junction. Simultaneously, a reversed bounced back spin current $J'_{s1}$ at the interface is generated, and transfer into $J_c$ due to inverse spin-Hall effect (ISHE), keeping the overall $J_c$ barely changed. Under this condition, an external magnetic field B is further applied along different directions, by which the magnetic moment in VGME is determined if B is large enough. When B is along Y-axis as shown in Fig. 5a, the magnetic moment in VGME $M_1$ is with the same direction as in the spin polarized electron, therefore, no spin moment is absorbed and $J_c$ is unchanged. When B is along X-axis as Fig. 5b illustrates, the magnetic moment in VGME $M_2$ is perpendicular to that of the spin polarized electron, the spin moment will be absorbed by VGME and the bounced back spin current decreases to $J'_{s2}$. Therefore, the overall charge current after ISHE is also lowered to $J'_c$. Such angular dependence of can be detected by measuring the change of longitudinal resistance $\Delta\rho_{xx}$ in Pt/VGME or Pt/a-C with respect to the rotation angle of B. Fig. 5c-e summarize the angular dependences of $\Delta\rho_{xx}$ of Pt/VGME and Pt/a-C on the rotation of magnetic field B in X-Y ($\alpha$ angle), Y-Z($\beta$ angle) and X-Z ($\gamma$ angle) planes, respectively. A periodical response of $\Delta\rho_{xx}$ with the angular change is clearly observed in Pt/VGME, whereas in Pt/a-C the resistance barely changes ($\Delta\rho_{xx}$ maintains near zero). It can be inferred that the angular dependent magnetoresistance is related to VGME film. The $\Delta\rho_{xx}$ along $\alpha$ angle shows a cosine feature, where the resistance is lower when B∥Y-axis, and higher when B⊥Y-axis, in accordance to the behavior of spin Hall magnetoresistance. We also notice that when B rotates in $\beta$ and $\gamma$ angle, same angular dependence appears with larger amplitude. This could be from the negative magnetoresistance of the VGME film, which is not our focus here. By subtracting $\Delta\rho_{xx}$ along $\gamma$ angle from $\Delta\rho_{xx}$ along $\beta$ angle, we obtained a change of $\Delta\rho_{xx}$ exactly equals to $\Delta\rho_{xx}$ along α angle. The angular dependent magnetoresistance behavior along α-angle implys a classic spin Hall magnetoresistance (SMR) process[30,31]. The observation of such SMR behavior at 300 K further proves the robust of the magnetization in VGME structure.

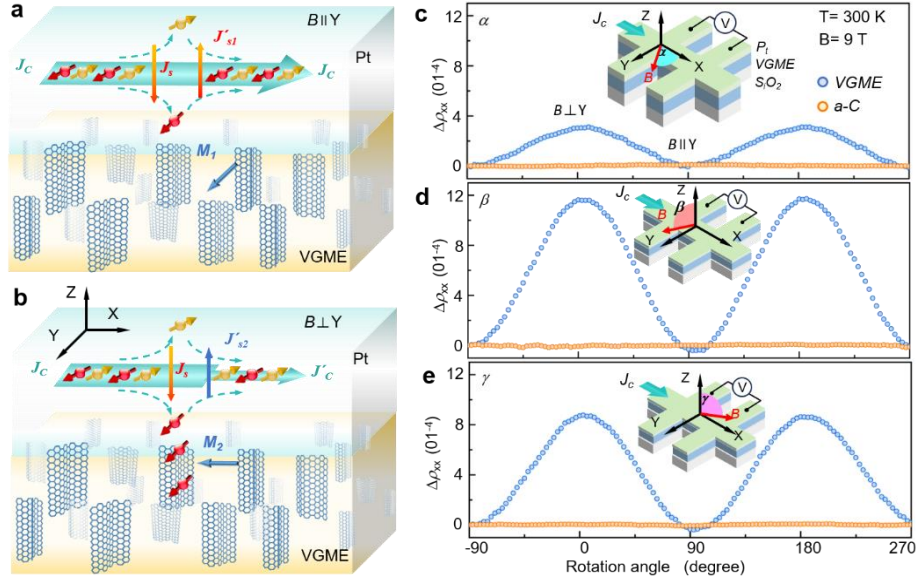

**Fig. 5 Spin Hall magnetoresistance in VGME structure. a-b,** The principle of current $J_c$ variation in response to external magnetic field B in parallel and perpendicular to Y axis. $J_s$ represents the spin current generated in Pt layer via spin Hall effect. $J'_{s1}$ and $J'_{s2}$ represent the bounced spin current when B||Y and B⊥Y, respectively. $M_1$ and $M_2$ represent the magnetization moment of VGME by B in different directions. **c-e,** the change of longitudinal resistance $\Delta\rho_{xx}$ in response to rotating B of 9 T around $\alpha$, $\beta$ and $\gamma$ angles at 300 K. The definition of each angle is illustrated inside each figure. Blue curves represent $\Delta\rho_{xx}$ of Pt/VGME bilayer, and orange curves represent $\Delta\rho_{xx}$ of Pt/a-C bilayer.

In conclusion, we produced an all-carbon film consisting of vertically aligned graphene edges, which wove into a three-dimensional hyperfine-porous network. A reproducible, full section of XMCD spectrum across C-1s K-edge absorption energy was detected for the first time, and the average amplitude of dichroism at 284.5 eV reaches ~0.005 at the 300 K, equals to unit magnetic moment of ~0.0006 $\mu_B$ per atom. The magnetization measured by XMCD is co-confirmed by hysteresis loop of M-H curve. The influence of magnetic impurities on magnetization was excluded by XANES and ICP-MS measurements. Spin transfer behavior of SMR was observed in the VGME film, which verified the long-range and robust features of the bulk ferromagnetic order at room temperature. Our work provides direct evidence of elementary resolved bulk carbon ferromagnetism at room temperature and clarifies its origin from π-electrons at graphene edges.

**Methods:**

**Sample preparation:**

A silicon wafer of 0.5 mm thickness with a 300 nm oxide layer was used as the substrate for the film growth. The thickness of the VGME was controlled via deposition time. For spin transport measurements, Pt layers with the thickness of 5 nm were sputtered via thermal evaporation.

**TEM characterization:**

For TEM characterization, the carbon film with 20nm thickness was transferred from Si substrate to a lacey carbon TEM grid for plan view characterization. The side view lamella was produced by a Focused Ion/Electron Dual beam system (Thermo-Fisher Scios). Both plan and side view characterization were performed by a double aberration-corrected electron microscope (Thermo-Fisher, Titan Themis G2) with Ceta2 CMOS camera and HAADF detector in TEM and HAADF STEM modes. The acceleration voltage of electron beam was set to 80kV to minimize the beam damage to the carbon film.

**Raman spectroscopy:**

For Raman measurement, the carbon film was deposited on Si substrate with the thickness of 20 nm. The spectra were obtained on HORIBA HR Evolution. An excitation laser with the wavelength of 532 nm was used with the emitting power of 1 $\mu$W, and the intensity coefficient of 3.2% was used for measurement. The numerical hole-diameter of the project lens was 1 $\mu$m. The measurement on each sample was repeated three times at different locations randomly selected on the film surface. The irradiating time of focused laser beam on each detecting area was less than 5 min.

**M-H curves:**

For M-H measurement, the carbon film was deposited on Si substrate with the thickness of 300 nm. The sample containing carbon film and $S_i$ substrate was cut into the in-plane size of 2×4 mm$^2$, and weighted on a digital weighing balance. The result was marked as $m_1$. The sample was glued onto a plastic sample holder in the position that the film surface was perpendicular to the external magnetic line. Hysteresis curve of the sample was obtained with a physical property measurement system (on Quantum Design, MPMS3). The raw magnetization result was recorded as $M_1$. The Si substrate without carbon film was measured with the same method. The weight and magnetization of the Si was recorded as $m_2$ and $M_2$. Since the thickness of the carbon film (300 nm) was much smaller than that of the substrate (0.5 mm), the contribution of the carbon sample to $m_2$ can be neglected (less than 1‰). Therefore, the magnetization response of carbon film can be calculated as $M_1 - m_1 \times M_2/m_2$.

**Laser ablation inductively coupled plasma mass spectrometry (LA-ICP-MS):**

For LA-ICP-MS measurements, the carbon film was deposited on Si substrate with the thickness of 300 nm. The as-deposited carbon film on Si substrate was cleaned by ultrasonic cleaning in acetone bath and dried naturally. The cleaned sample was mounted flatly in the sample chamber of laser-ablation chamber (NWR193HE) which was coupled with the ICP-MS system. The chamber was closed and Ar was introduced as background. A pulse laser with the frequency of 5 Hz was introduced to evaporate the film. The sampling time at each position was 40 s, and the process repeated at 24 different positions in total. A NIST-610 sample was used for concentration calibration of the most possible metallic impurities, including Al, P, Ti, V, Cr, Mn, Fe, Co, Ni, Cu, Zn, and Pb.

**X-ray photon spectroscopy(XPS):**

For XPS measurement, the carbon film was deposited on Si substrate with the thickness of 20 nm. The as deposited sample was cleaned in acetone bath and dried naturally before measurement. The film surface was tested firstly, followed by Ar plasma etching for 2 min to remove the part that contacted with atmosphere, and the same test was carried out on the fresh-exposed surface. An X-ray source was used, and the whole spectrum were collected in the energy range from 0 to 1,350 eV. O 1s and C 1s spectra was magnified to check the composition difference between film body and surface.

**X-ray magnetic circular dichroism**

For XMCD measurement, 10 times of repeatedly scan were carried out continuously. Each time a MXCD curve was derived from a pair of XANES spectra taken under opposite magnetic fields. The integrated area of XMCD curve $A_{XMCD}$ was calculated by summing up the normalized amplitude of all data points. The area of XANES spectra $A_{XANES}$ was derived with the same process. The ratio of the area of XMCD to the area of XANES was calculated, and the value $r = \frac{A_{XMCD}}{A_{XANES}}$ is considered as the effective magnetic moment from each carbon atom.

**Device fabrication:**

For transport measurement, the carbon film was deposited on Si substrate with the thickness of 20 nm. The sample was cut into pieces (in-plane size 4 × 4 mm$^2$) with a diamond pencil, followed by 3 min ultrasonic cleaning in acetone bath. For Hall measurements, a Hall-bar shaped photoresist coating was fabricated on the carbon film by lithography method, and Ar etching was introduced to remove the photoresist and the uncovered carbon film, what left was a Hall-bar shaped carbon film. The line width of the Hall bar is 100 $\mu$m (see Fig. S4). The electrodes were connected to a standard sample puck with Pt wires (diameter of 20 $\mu$m) by using a wiring bonding machine. For spin transport measurements, a Pt film was sputtered on the whole surface of the carbon film by DC sputtering in Ar atmosphere, followed by Hall-bar fabrication process as for Hall measurements, with the Ar etching time slightly longer.

We also prepared the devices by directly attaching Pt wire on carbon and Pt/carbon film surfaces with silver paste instead of fabricating the Hall-bar sample. The results showed that the influence of different techniques on measurement results can be neglected.

**Data availability:**

The data that support the findings of this study are available from the corresponding author on reasonable request.

**References:**

1. Son, Y.-W., Cohen, M. L. & Louie, S. G. Half-metallic graphene nanoribbons. *Nature* **444**, 347–349 (2006).

2. Magda, G. Z. *et al.* Room-temperature magnetic order on zigzag edges of narrow graphene nanoribbons. *Nature* **514**, 608–611 (2014).

3. Blackwell, R. E. *et al.* Spin splitting of dopant edge state in magnetic zigzag graphene nanoribbons. *Nature* **600**, 647–652 (2021).

4. Mishra, S. *et al.* Collective All-Carbon Magnetism in Triangulene Dimers**. *Angew. Chem. Int. Ed.* **59**, 12041–12047 (2020).

5. Mishra, S. *et al.* Topological frustration induces unconventional magnetism in a nanographene. *Nat. Nanotechnol.* **15**, 22–28 (2020).

6. Pizzochero, M. *et al.* Edge Disorder in Bottom-Up Zigzag Graphene Nanoribbons: Implications for Magnetism and Quantum Electronic Transport. *J. Phys. Chem. Lett.* **12**, 4692–4696 (2021).

7. Friedrich, N. *et al.* Magnetism of Topological Boundary States Induced by Boron Substitution in Graphene Nanoribbons. *Phys. Rev. Lett.* **125**, 146801 (2020).

8. Slota, M. *et al.* Magnetic edge states and coherent manipulation of graphene nanoribbons. *Nature* **557**, 691–695 (2018).

9. Rizzo, D. J. *et al.* Inducing metallicity in graphene nanoribbons via zero-mode superlattices. *Science* **369**, 1597–1603 (2020).

10. Tschirhart, C. L. *et al.* Imaging orbital ferromagnetism in a moiré Chern insulator. *Science* **372**, 1323–1327 (2021).

11. Sharpe, A. L. *et al.* Emergent ferromagnetism near three-quarters filling in twisted bilayer graphene.

*Science* **365**, 605–608 (2019).

12. Esquinazi, P. *et al.* Induced Magnetic Ordering by Proton Irradiation in Graphite. *Phys. Rev. Lett.* **91**, 227201 (2003).

13. Yazyev, O. V. Magnetism in Disordered Graphene and Irradiated Graphite. *Phys. Rev. Lett.* **101**, 037203 (2008).

14. Chen, J.-H., Li, L., Cullen, W. G., Williams, E. D. & Fuhrer, M. S. Tunable Kondo effect in graphene with defects. *Nat. Phys.* **7**, 535–538 (2011).

15. Ohldag, H. *et al.* π-Electron Ferromagnetism in Metal-Free Carbon Probed by Soft X-Ray Dichroism. *Phys. Rev. Lett.* **98**, 187204 (2007).

16. Zhang, Y., Gao, F., Gao, S. & He, L. Tunable magnetism of a single-carbon vacancy in graphene. *Sci. Bull.* **65**, 194–200 (2020).

17. Nair, R. R. *et al.* Dual origin of defect magnetism in graphene and its reversible switching by molecular doping. *Nat. Commun.* **4**, 2010 (2013).

18. Hu, W. *et al.* Embedding atomic cobalt into graphene lattices to activate room-temperature ferromagnetism. *Nat. Commun.* **12**, 1854 (2021).

19. Zhang, X., Qi, S. & Xu, X. Long-range and strong ferromagnetic graphene by compensated n–p codoping and π–π stacking. *Carbon* **95**, 65–71 (2015).

20. Ugartemendia, A., Garcia−Lekue, A. & Jimenez−Izal, E. Tailoring magnetism in silicon-doped zigzag graphene edges. *Sci. Rep.* **12**, 13032 (2022).

21. Park, J. M., Cao, Y., Watanabe, K., Taniguchi, T. & Jarillo-Herrero, P. Tunable strongly coupled superconductivity in magic-angle twisted trilayer graphene. *Nature* **590**, 249–255 (2021).

22. Wang, Y. *et al.* Room-Temperature Ferromagnetism of Graphene. *Nano Lett.* **9**, 220–224 (2009).

23. Červenka, J., Katsnelson, M. I. & Flipse, C. F. J. Room-temperature ferromagnetism in graphite driven by two-dimensional networks of point defects. *Nat. Phys.* **5**, 840–844 (2009).

24. Sepioni, M. *et al.* Limits on Intrinsic Magnetism in Graphene. *Phys. Rev. Lett.* **105**, 207205 (2010).

25. Nair, R. R. *et al.* Spin-half paramagnetism in graphene induced by point defects. *Nat. Phys.* **8**, 199–202 (2012).

26. Wang, C., Diao, D., Fan, X. & Chen, C. Graphene sheets embedded carbon film prepared by electron irradiation in electron cyclotron resonance plasma. *Appl. Phys. Lett.* **100**, 231909 (2012).

27. Wang, C., Zhang, X. & Diao, D. Nanosized graphene crystallite induced strong magnetism in pure carbon films. *Nanoscale* **7**, 4475–4481 (2015).

28. Chen, C. T. *et al.* Experimental Confirmation of the X-Ray Magnetic Circular Dichroism Sum Rules for Iron and Cobalt. *Phys. Rev. Lett.* **75**, 152–155 (1995).

29. N'Diaye, A. *et al.* Interplay between Transition-Metal K-edge XMCD and Magnetism in Prussian Blue Analogs. *ACS Omega* **7**, 36366–36378 (2022).

30. Avci, C. O. *et al.* Unidirectional spin Hall magnetoresistance in ferromagnet/normal metal bilayers. *Nat. Phys.* **11**, 570–575 (2015).

31. Oyanagi, K. *et al.* Paramagnetic spin Hall magnetoresistance. *Phys. Rev. B* **104**, 134428 (2021).

**Acknowledgements:**

The authors would like to thank Prof. Kai Chen for discussion on *XMCD* technique, and thank Dr. Duo Zhao for discussion on *SMR* measurement. We also thank BL12B beamline of National Synchrotron Radiation Laboratory (NSRL, China) and BL07U beamline of Shanghai Synchrotron Radiation Facility (SSRF). The work is supported by National Natural Science Foundation of China (grant No. 51875364)

**Author contributions:**

C.W, D.D and E.S. designed the research. N.J., M.Y. and X. Z. performed the transmission electron microscopy analysis. Z.Y. and X. M. prepared the sample and assisted with all measurements. T.K. and S. D.

designed the M-H and SMR measurements, Q. L, W. Y. and D. H. assisted with *XMCD* measurements. L.Y. assisted with M-H measurements. D. D. and E.J. supervised the research. C.W. and D.D. co-wrote the paper. All authors discussed the results and commented on the manuscript.

**Competing interests:**

The authors declare no competing interests.